# The test of investors' behavioral bias through the price discovery process in cryptoasset exchange: Transactional-Level Evidence from Thailand

Roongkiat Ratanabanchuen[b+], Kanis Saengchote [b++], Voraprapa Nakavachara[a],

Thitiphong Amonthumniyom[c], Pongsathon Parinyavuttichai[c], Polpatt Vinaibodee[c]

Abstract:

Analyzing investors' trading behavior in cryptoasset markets provides new evidence supporting the theory that retail investors likely exhibit behavioral biases. We investigate the price discovery process between Thailand's most highly liquid exchange and the global exchange. Under the no-arbitrage assumption, bid-offer quotes in the local exchange should quickly move to match the new price levels in the global exchange, as the price process of cryptoassets in the local exchange does not contain new information that can lead the price dynamic in the global exchange. We analyze intraday bid-offer quotes and investors' portfolio positions and find that investors exhibit the disposition effect by attempting to sell their profitable positions during market upturns. The rate of bid-offer movement is significantly slow to match a new global price level only in situations when most investors in the market are in profit. These insights are crucial as they suggest that the risk-return characteristics of asset prices between the bull and bear market may differ, resulting from investors' behavioral biases.

---

[a] Faculty of Economics, Chulalongkorn University

[b] Chulalongkorn Business School, Chulalongkorn University

[c] The Securities and Exchange Commission of Thailand

[+] First Author: roongkiat@cbs.chula.ac.th

[++] Corresponding Author: kanis@cbs.chula.ac.th

## Introduction

In the rapidly evolving financial markets, cryptoasset trading stands out as a distinctive and fertile ground for exploring investor behavior, diverging significantly from the well-trodden paths of traditional asset classes like stocks. The inherent accessibility of cryptoassets, coupled with their high volatility and relative nascency, presents a unique set of challenges and opportunities that profoundly influence investor decision-making. Unlike traditional markets, where decades of empirical research have shed light on investor behaviors, cryptoasset remains a nascent field ripe for exploration.

This study leverages the rich yet largely untapped dataset of cryptoasset trading activities to delve into how cognitive biases manifest in an environment characterized by less regulation, greater price swings, and higher uncertainty. Our collaboration with the Securities and Exchange Commission of Thailand provides an unprecedented opportunity to examine the intricacies of investor behaviors in the cryptoasset market, utilizing comprehensive order books and investor portfolio positions from a leading local exchange during the pivotal years of 2021-2022.

Through this lens, we aim to unravel the complex interplay between global market movements and localized trading behaviors in the crypto domain. This analysis is particularly salient given the distinctive features of cryptoasset markets, such as the 24/7 trading cycle, the diversity of participants, and the influence of global events on local trading dynamics. By focusing on cryptoassets, we contribute to a broader understanding of financial market behaviors under novel conditions and offer insights into the adaptive strategies of investors in a market environment that defies traditional finance paradigms.

Conventional economic theory posits that, under the no-arbitrage assumption, local market bids and offers should rapidly converge to global price levels. However, our observations of lagged responses in these bids and offers across varying market conditions suggest the presence of behavioral biases. This deviation from theoretical expectations is particularly intriguing in cryptocurrency markets, which are known for their volatility and the predominance of retail investors—a demographic often susceptible to behavioral anomalies.

Building upon Campbell's (2000) critique that "[Behavioral models] cannot be tested using aggregate consumption or the market portfolio because rational utility-maximizing investors neither consume aggregate consumption (some is accounted for by nonstandard investors) nor hold the market portfolio (instead they shift in and out of the stock market)" (p. 1551), our study employs intraday bid-offer quotes to examine these biases at high frequency. This approach enables a nuanced understanding of retail investor behavior under diverse market conditions by analyzing trading volume, bid-offer spreads, and quote movements.

Our study focuses on the three most liquid cryptoassets— Bitcoin (BTC), Ether (ETH), and Ripple (XRP)—utilizing 30-minute interval limit order book data from Thailand's largest cryptoasset exchange by trading volume. This exchange, commanding a market share of approximately 87% and an average daily trading volume of about 2.7 billion baht (approximately 75 million US dollars), offers a robust platform for our analysis.

The choice of the Thai cryptoasset market for this study is particularly compelling due to several factors. Firstly, the market predominantly comprises retail investors who are more prone to behavioral biases. Secondly, the nature of order execution in this market, primarily driven by price movements in global markets rather than new, local information, provides a clear window into behavioral responses to external stimuli. Thirdly, the wider bid-offer spreads relative to global markets allow for a distinct analysis of retail investor orders rather than market-maker activity. Lastly, the global interconnectedness of cryptoasset prices presents an excellent opportunity to investigate the efficiency of price discovery processes and the persistence of arbitrage opportunities, which can further illuminate investors' behavioral biases.

Our empirical analysis reveals two distinct stages of price adjustment in the local market in response to global price movements. The initial 30-minute period shows a 95% adjustment to global prices, which then increases to nearly 97% in the subsequent half-hour, with BTC exhibiting the most rapid adjustment. This pattern indicates investor behavior under different market conditions and provides insights into the prevalent disposition effect in the local market. Specifically, our findings suggest that investors are inclined to sell profitable positions during price rises, as evidenced by the local offer quotes that consistently undercut global price levels during market uptrends. Conversely, the sluggish response of bid quotes to global price increases points to a risk-averse attitude among investors when in profit, potentially reflecting biases like conservativeness and anchoring.

The results establish a connection between the behavioral biases of retail investors and market price responses, presenting evidence that transcends mere noise trading. We critically examine several behavioral finance theories, including prospect theory (Kahneman and Tversky, 1979), the house-money effect (Thaler and Johnson, 1990), and attribution biases (Daniel et al., 1998; Gervais and Odean, 2001) and rigorously test the null hypothesis of rationality against these biases.

This paper is structured as follows: Section I delineates the characteristics of cryptoasset markets. Section II delves into the statistical test of price discovery, while Section III proposes the research methodology that uses the price discovery process to test investor biases. Section IV presents our empirical findings on bid-offer spreads and the price discovery process for BTC, ETH, and XRP in the Thai market. Finally, Section V summarizes our key insights and conclusions.

## I. Background on the Thai cryptoasset market and data used in this study.

Cryptoassets are an alternative asset class that attracts a particular type of investor. For Thailand, ranked 8$^{th}$ in the 2022 Global Crypto Adoption Index by Chainanalysis,[1] over 70% of investors are 18 to 24.[2] The Thai market starkly contrasts the US, where Aiello et al. (2023) demonstrate that cryptoasset investors come from all income levels, and households appear to treat cryptoassets like traditional assets.

During our analysis, five cryptoasset exchanges were approved by Thailand's Securities and Exchange Commission (SEC). As of December 2022, there were around 2.9 million accounts in all five cryptoasset exchanges, which is relatively high compared to 5.14 million stock trading accounts in the Stock Exchange of Thailand. Across the five exchanges, 191 cryptoassets are available for trade.

[1] https://www.chainalysis.com/blog/2022-global-crypto-adoption-index/
[2] https://www.statista.com/statistics/1294429/thailand-distribution-of-cryptocurrency-investors-by-age-group/

The trading volume of the largest exchange accounted for nearly 90 % of the total daily trading volume, which stood at around 3.23 billion baht per day (approximately 92 million US dollars). The exchange grew rapidly, from just 60,000 accounts at the beginning of 2020 to almost one million at the end of 2022, where around 20-30 % of the accounts traded at least once a week.

Despite the growing global acceptance of cryptoasset investment, the integration of these assets into the portfolios of institutional investors remains nascent. Consequently, retail investors constitute approximately 84 % of trading activities in the authorized cryptoasset exchange examined in this study, making the market a fitting avenue for investigating behavioral biases among retail investors. The specifics of the local exchange used in this analysis are detailed in Table 1.

**Table 1:** Key characteristics of the local exchange in Thailand

| | 2020 | 2021 | 2022 |
|---|---|---|---|
| Number of listed cryptoassets | 54 | 94 | 109 |
| Number of trading accounts | 60,340 | 924,601 | 989,865 |
| Number of domestic accounts | 56,752 | 915,125 | 979,499 |
| Number of foreign accounts | 3,602 | 9,491 | 10,440 |
| Trading value per day (unit: million baht) | | | |
| Mean | 565.18 | 4,581.53 | 2,751.80 |
| Standard deviation | 309.62 | 2,958,46 | 1,777.56 |
| Min | 112.64 | 888.61 | 257.98 |
| Max | 1,580.40 | 25,542.89 | 13,225.26 |
| Trading value per account per day (unit: baht) | | | |
| Mean | 12,464.18 | 5,895.77 | 5,528.39 |
| Standard deviation | 52,428.84 | 30,551.93 | 28,521.39 |
| Min | 0.01 | 0.01 | 0.01 |
| Max | 7,549,668.99 | 20,517,059.20 | 14,962,500.11 |
| Proportion of retail investors | | | |
| % of the number of accounts | 99.97 | 99.99 | 99.98 |
| % of trading value | 91.74 | 94.28 | 84.06 |
| Proportion of juristic investors | | | |
| % of the number of accounts | 0.03 | 0.01 | 0.02 |
| % of trading value | 8.26 | 5.72 | 15.94 |
| Market share of trading volume (unit %) | 88.92 | 80.59 | 87.07 |

Cryptoasset markets operate without regulatory constraints on bid-offer spread limits, minimum tick sizes, or caps on daily price fluctuations. This feature allows for endogenous and continuous adjustment of spreads and prices. Cryptoasset exchanges are entirely digital, with liquidity self-generated by investors who place buy or sell limit orders at predetermined prices. These orders are aggregated in the exchange’s order book.

The electronic trading mechanism in cryptoasset exchanges parallels a continuous auction system, where trades are automatically executed upon the submission of market orders. The market price, therefore, is a culmination of the evolution of bid-offer quotes and the timing of order execution. The observable bid-offer quotes represent the extant public information, while executed trades reflect the actions of investors acting on private information, as outlined by Glosten and Milgrom (1985) and Kyle (1985).

Unlike traditional assets like stocks and bonds, which are typically confined to specific exchanges due to regulatory and logistical frameworks, cryptoassets are characterized by their ability to be traded across a myriad of global exchanges. This unique feature stems from the decentralized nature of cryptoassets and the absence of central authority oversight. The global network of crypto exchanges facilitates diverse trading opportunities and provides a broader scope for price discovery and arbitrage. However, it also introduces complexities in tracking and understanding market dynamics, underscoring the significance of our study in deciphering investor behaviors within this multifaceted landscape.

Our study utilizes anonymized, transaction-level data from participants trading cryptocurrencies on Thailand's largest licensed digital asset exchange, which operates under Thai regulations. Following the Notification of the Securities and Exchange Commission No. GorThor. 26/2562, these exchanges are required to report customers' trading activities to the SEC from November 2020. Thus, our dataset spans from November 2020 to December 2022. Data management and handling were conducted by authors affiliated with the SEC, adhering strictly to the SEC's data protection protocol.

The dataset encompasses detailed information on buy and sell transactions, including the date and time of trade execution, cryptocurrency type, execution prices, trading volume, pseudonymized customer identifiers, and account type. Using limit order book data, we construct a 30-minute bid and offer prices for Bitcoin (BTC), Ether (ETH), and Ripple (XRP), the main cryptoassets in Thailand and globally.

## II. Price discovery analysis

This study investigates price discovery in cryptoasset markets, a topic that has attracted increasing attention in recent financial literature. While previous research, such as Brandvold et al. (2015), Dimpfl and Peter (2019), and Urquhart (2016), has predominantly focused on identifying leading markets in global cryptoasset price formation, we employ a different approach.[3] We employ a lead-lag analysis between a prominent local exchange in Thailand and Kraken (representing international market price), not to ascertain market leadership in price formation but to gauge the efficiency of price discovery in the local market and to interpret investor behavior in varying market conditions.

Price discovery, as illustrated by Hasbrouck (1995, 2003), Booth et al. (2002), and Huang (2002), refers to the process through which markets assimilate new information and adjust prices accordingly. The crux of this concept lies in identifying the market that leads in integrating new information, thereby influencing prices in other markets trading the same asset. Early studies in this domain, such as Hasbrouck (1995) and Gonzalo and Granger (1995), have predominantly focused on equities, examining information shares and the error correction process in price movements across different stock exchanges.

In the context of cryptoassets, the principle of no arbitrage should theoretically ensure the alignment of prices across different exchanges. This expectation implies that price series should be cointegrated. The individual price series should be integrated of order 1, denoted as

[3] Brandvold et al. (2015) shows that Mt. Gox is the leading market for bitcoin spot prices where prices of bitcoin in other exchanges will follow. Dimpfl and Peter (2019) incorporate microstructure noises using the methodology proposed by Putnins (2013) and find that trading in Bitfinex lead prices in Kraken and Poloniex. In a related study, Urquhart (2016) employs statistical tests to prove that cryptoasset markets across the globe appear to be more efficient over time.

I(1), meaning the price series are nonstationary in levels but exhibit stationarity in their differences, adhering to a long-term equilibrium. If the price series across different markets of the same asset are cointegrated, they share one or more common stochastic factors.

It is crucial to distinguish cointegration from mere correlation; while the latter measures the degree of synchronous movement between price series, cointegration tests for lead-lag relationships and ensures that prices across markets do not drift apart due to common stochastic trends. Moreover, cointegration also refers to the statistical property that price series are nonstationary, but the cointegration terms of those price series are stationary. In other words, the price series of the same assets in different markets follow the same long-run equilibrium.

Our focus on cryptoassets, specifically BTC, entails examining whether the price series in the local Thai exchange and a global exchange like Kraken share a common factor—what we term the 'implicit efficient price.' This efficient price, driven by fundamental information, dictates the long-term price level changes across markets. Our assumption that cryptoasset prices in different markets conform to this common implicit efficient price aligns with the economic rationale of eliminating arbitrage opportunities.

The cornerstone of our analysis is the vector error correction model (VECM), a statistical tool employed to decipher the cointegration relationships between cryptoasset prices in a Thai exchange and the Kraken exchange. The VECM allows for a nuanced examination of the price formation patterns, particularly in response to price movements in international cryptoasset markets. The methodology implemented here hinges on a 'lead-lag' return regression approach. It involves regressing one return series against the leading and lagging intervals of another.

To illustrate the VECM, suppose that $P_{A,t}$ is the price of BTC in Market A and $P_{B,t}$ is the price of BTC in Market B, and the two prices are cointegrated. Then, the cointegration equation can be written as:

$$P_{A,t} = \beta_0 + \beta_1 P_{B,t} + \eta_t$$

$$\eta_t = P_{A,t} - \beta_1 P_{B,t} - \beta_0$$

Both $P_{A,t}$ and $P_{B,t}$ should be I(1). Thus, the price process could be expressed by an autoregressive process, where i can be for market A or B.

$$P_{i,t} = \phi_{i0} + \phi_{i1} P_{i,t-1} + \nu_{i,t}$$

With recursive substitution and the cointegration equation, it can be shown that the differenced price series can be written as:

$$\Delta P_{A,t} = \alpha_A \eta_{t-1} + \sum_{k=1}^{p} \gamma_A \Delta P_{A,t-k} + \sum_{k=1}^{p} \delta_A \Delta P_{B,t-k} + \epsilon_{A,t}$$

$$\Delta P_{B,t} = \alpha_B \eta_{t-1} + \sum_{k=1}^{p} \gamma_B \Delta P_{A,t-k} + \sum_{k=1}^{p} \delta_B \Delta P_{B,t-k} + \epsilon_{B,t}$$

Where $\eta_{t-1}$ is the error correction term, and p is the lag. One key aspect of our model is the flexibility in choosing the lag periods, which range from seconds to minutes, offering a granular view of price dynamics. The coefficient estimations derived from the VECM are instrumental in uncovering the patterns of price discovery and the interplay between local and global markets. Detailed order book data allows us to construct high-frequency bid and offer prices to estimate the VECM separately for bids and offers.

We focus on the best bid and offer prices within 30-minute intervals between November 2020 and December 2022. To align the data from the global exchange with the local context, we convert cryptoasset values into the local currency using the prevailing exchange rates for each timeframe.

Contrary to utilizing the midpoint of quotes from local and global exchanges—a common approach in many empirical microstructure studies—we opt for a more nuanced method. The rationale behind this choice is supported by findings from Cao, Hansch, and Wang (2009) and Pascual and Veredas (2010), which suggest that using the midpoint can lead to a loss of critical information, particularly in assessing investors' buying and selling intentions. In granular timeframes, analyzing separate bid and offer quotes reveals insightful dynamics and aids in testing investors' behavioral biases. The coefficients derived from these leads and lags for the whole period of study, along with the error correction terms, are presented in Table 2

Our analysis diverges from the traditional approach of estimating key parameters that Hasbrouck (1995) and Gonzalo and Granger (1995) proposed. Instead, we focus on the VECM's coefficients to interpret the price discovery process. Our findings align with the expectation that Thai exchange prices follow global market trends, a hypothesis corroborated by the VECM estimations.

A critical observation from our model is the statistically and economically significant positive coefficients for the second and third lags of price differences in the global exchange. When modeling the price differences in the local exchange as the dependent variable, this result suggests that global exchange prices (across all three analyzed cryptoassets) precede local exchange prices by 1-2 lags. This delay indicates the time for local exchange prices to align with new global price levels.

Conversely, the coefficients corresponding to all lags of price differences in the local exchange are statistically insignificant when the global exchange price differences are the dependent variable. This finding underscores the absence of new information on local exchange prices for price discovery. It corroborates the view that retail investors in the Thai market primarily engage in trading activities that mirror global market movements rather than being a source of information.

**Table 2:** VECM estimations between the Thai exchange and Kraken

Panel A: Bitcoin (BTC)

| | Constant term | Cointegration term | Price differences in the Thai exchange | | | Price differences in Kraken | | |
|---|---|---|---|---|---|---|---|---|
| | | First lag | First lag | Second lag | Third lag | First lag | Second lag | Third lag |
| **BTC bid quote** | | | | | | | | |
| Dependent variable: price differences in the Thai exchange | | | | | | | | |
| Coefficient | -0.0482 | -0.0333*** | -0.5328*** | -0.2880*** | -0.0743*** | 0.6775*** | 0.5797*** | 0.2940*** |
| Standard errors | 34.9104 | 0.0017 | 0.0051 | 0.0051 | 0.0038 | 0.0050 | 0.0060 | 0.0063 |
| Dependent variable: price difference in Kraken | | | | | | | | |
| Coefficient | 0.2930 | -0.0055*** | -0.0017 | -0.0048 | -0.0109*** | -0.0272*** | -0.0167 | 0.0020 |
| Standard errors | 38.7371 | 0.0019 | 0.0057 | 0.0056 | 0.0042 | 0.0056 | 0.0067 | 0.0070 |
| Cointegrating equations | | | | | | | | |
| | Constant term | Price in the Thai exchange | Price in Kraken | | | | | |
| Coefficient | 19,785.34 | 1.0000 | -1.0252*** | | | | | |
| Standard errors | | | 0.0024 | | | | | |
| **BTC offer quote** | | | | | | | | |
| Dependent variable: price differences in the Thai exchange | | | | | | | | |
| Coefficient | -0.0573 | -0.0317*** | -0.5324*** | -0.2950*** | -0.0715*** | 0.6655*** | 0.5426*** | 0.2842*** |
| Standard errors | 34.4686 | 0.0016 | 0.0051 | 0.0051 | 0.0038 | 0.0049 | 0.0059 | 0.0061 |
| Dependent variable: price difference in Kraken | | | | | | | | |
| Coefficient | 0.4003 | -0.0045** | -0.0192*** | -0.0068 | -0.0002 | -0.0408*** | 0.0010 | 0.0069 |
| Standard errors | 39.0102 | 0.0018 | 0.0058 | 0.0058 | 0.0043 | 0.0056 | 0.0067 | 0.0070 |
| Cointegrating equations | | | | | | | | |
| | Constant term | Price in the Thai exchange | Price in Kraken | | | | | |
| Coefficient | 21,306.45 | 1.00 | - 1.0300*** | | | | | |
| Standard errors | | | 0.0025 | | | | | |

*** p<0.01, ** p<0.05, * p<0.1

Panel B: Ether (ETH)

| | Constant term | Cointegration term | Price differences in the Thai exchange | | | Price differences in Kraken | | |
|---|---|---|---|---|---|---|---|---|
| | | First lag | First lag | Second lag | Third lag | First lag | Second lag | Third lag |
| **ETH bid quote** | | | | | | | | |
| Dependent variable: price differences in the Thai exchange | | | | | | | | |
| Coefficient | -0.0011 | -0.0881*** | -0.3763*** | -0.1701*** | -0.0497*** | 0.3746*** | 0.3349*** | 0.1467*** |
| Standard errors | 3.3930 | 0.0025 | 0.0053 | 0.0052 | 0.0045 | 0.0052 | 0.0056 | 0.0054 |
| Dependent variable: price difference in Kraken | | | | | | | | |
| Coefficient | 0.8569 | -0.0001 | 0.0168*** | 0.0087 | -0.0057 | -0.2306*** | -0.0605*** | -0.0218*** |
| Standard errors | 3.8519 | 0.0029 | 0.0060 | 0.0059 | 0.0051 | 0.0059 | 0.0064 | 0.0061 |
| Cointegrating equations | | | | | | | | |
| | Constant term | Price in the Thai exchange | Price in Kraken | | | | | |
| Coefficient | 703.15 | 1.00 | -1.0169*** | | | | | |
| Standard errors | | | 0.0011 | | | | | |
| **ETH offer quote** | | | | | | | | |
| Dependent variable: price differences in the Thai exchange | | | | | | | | |
| Coefficient | -0.0191 | -0.0740*** | -0.4195*** | -0.1990*** | -0.0690*** | 0.5378*** | 0.3645*** | 0.1678*** |
| Standard errors | 3.3045 | 0.0023 | 0.0052 | 0.0053 | 0.0043 | 0.0062 | 0.0068 | 0.0068 |
| Dependent variable: price difference in Kraken | | | | | | | | |
| Coefficient | 0.6827 | -0.0021 | -0.0062 | 0.0037 | -0.0130*** | -0.0187*** | -0.0039 | -0.0001 |
| Standard errors | 3.0185 | 0.0021 | 0.0048 | 0.0048 | 0.0040 | 0.0057 | 0.0062 | 0.0063 |
| Cointegrating equations | | | | | | | | |
| | Constant term | Price in the Thai exchange | Price in Kraken | | | | | |
| Coefficient | 618.90 | 1.00 | -1.0209*** | | | | | |
| Standard errors | | | 0.0013 | | | | | |

*** p<0.01, ** p<0.05, * p<0.1

Panel C: Ripple (XRP)

| | Constant term | Cointegration term | Price differences in the Thai exchange | | | Price differences in Kraken | | |
|---|---|---|---|---|---|---|---|---|
| | | First lag | First lag | Second lag | Third lag | First lag | Second lag | Third lag |
| **XRP bid quote** | | | | | | | | |
| Dependent variable: price differences in the Thai exchange | | | | | | | | |
| Coefficient | 0.0000 | -0.0855*** | -0.4885*** | -0.2349*** | -0.0768*** | 0.6422*** | 0.5206*** | 0.2326*** |
| Standard errors | 0.0014 | 0.0027 | 0.0053 | 0.0052 | 0.0039 | 0.0059 | 0.0067 | 0.0068 |
| Dependent variable: price difference in Kraken | | | | | | | | |
| Coefficient | -0.0001 | -0.0085*** | -0.0002 | -0.0097* | -0.0208*** | -0.0451*** | -0.0393*** | -0.0051 |
| Standard errors | 0.0014 | 0.0027 | 0.0053 | 0.0052 | 0.0039 | 0.0060 | 0.0067 | 0.0069 |
| Cointegrating equations | | | | | | | | |
| | Constant term | Price in the Thai exchange | Price in Kraken | | | | | |
| Coefficient | 0.45 | 1.00 | -1.0282*** | | | | | |
| Standard errors | | | 0.0015 | | | | | |
| **XRP offer quote** | | | | | | | | |
| Dependent variable: price differences in the Thai exchange | | | | | | | | |
| Coefficient | 0.0000 | -0.1111*** | -0.5089*** | -0.2716*** | -0.0955*** | 0.5511*** | 0.4972*** | 0.2704*** |
| Standard errors | 0.0018 | 0.0032 | 0.0054 | 0.0054 | 0.0043 | 0.0074 | 0.0079 | 0.0080 |
| Dependent variable: price difference in Kraken | | | | | | | | |
| Coefficient | -0.0001 | -0.0050** | -0.0044 | -0.0109*** | -0.0018 | -0.0434*** | -0.0351*** | -0.0038 |
| Standard errors | 0.0014 | 0.0025 | 0.0042 | 0.0042 | 0.0034 | 0.0058 | 0.0062 | 0.0063 |
| Cointegrating equations | | | | | | | | |
| | Constant term | Price in the Thai exchange | Price in Kraken | | | | | |
| Coefficient | 0.4433 | 1.00 | -1.0358*** | | | | | |
| Standard errors | | | 0.0015 | | | | | |

*** p<0.01, ** p<0.05, * p<0.1

## III. Motivating the relationship between price discovery and behavioral biases

The estimated coefficients shown in Table 2 can be used to formulate the price impulse response function. This function quantifies the rate at which local exchange prices respond to price impulses from the global exchange. To aid in interpreting price impulses, we define 'relative quote value,' represented by the following equation:

$$Relative\ Quote\ Value_t = \frac{Quote\ in\ the\ Local\ Exchange_t}{Price\ Impulse\ Level\ in\ the\ Global\ Exchange_0}$$

The price level from the impulse response function is expressed as a function of time from a 'shock' to the system of equations. If the price response is efficient, the relative quote value should immediately be 100%. We will use this metric in our further investigation of investor behavioral biases.

As categorized by Barberis and Thaler (2003), behavioral finance theories can be broadly divided into two types: biases stemming from beliefs and biases resulting from preferences. Belief-based biases arise from individuals' misinterpretation of information. For instance, self-attribution bias, as described by Daniel et al. (1998) and Gervais and Odean (2001), manifests in individuals taking credit for successes while attributing failures to external factors. In the context of cryptoasset markets, if investors exhibit this self-attribution bias, we expect them to gain confidence from profitable trades, leading them to take on greater risks when cryptoasset prices rise. Consequently, the bid response should align more rapidly with new global price levels during market upturns compared to downturns. This hypothesis is visually represented in Figure 1, where we anticipate a positive correlation between relative quote value and portfolio returns.

**Figure 1:** The relationship between the relative quote value and market returns (or portfolio returns) according to different behavioral bias assumptions

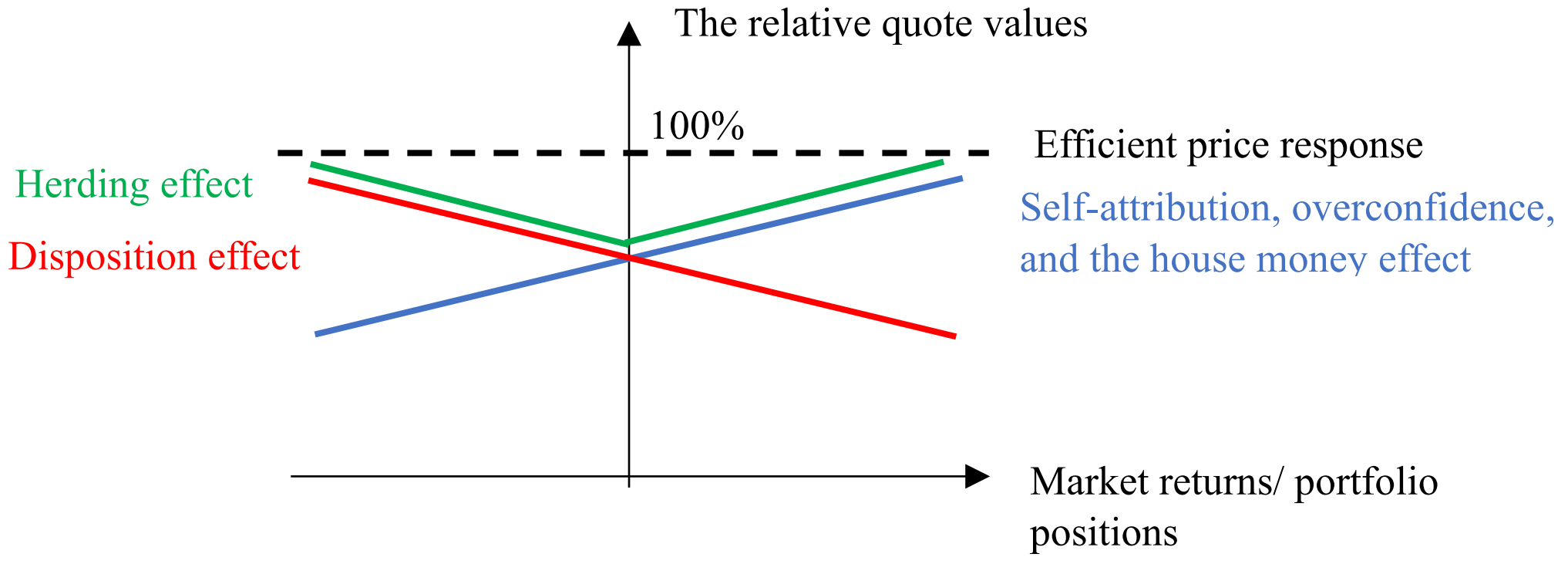

Regarding biases resulting from preferences, investor behavior often deviates from the rational utility-maximization model. This deviation is primarily grounded in the prospect theory, which posits that individuals' utility functions are convex under losses and concave under gains, a departure from the rationality suggested by the Von Neumann–Morgenstern axioms. According to this theory, investors exhibit risk-seeking behavior when they experience losses. Kahneman and Tversky (1979) encapsulate this tendency, stating, "[A] person who has not made peace with his losses is likely to accept gambles that would be unacceptable to him otherwise" (p. 287). Consequently, this behavioral propensity indicates that investors might engage more aggressively in trading during periods of pronounced negative price movements than positive ones.

Furthermore, the prospect theory underpins the 'disposition effect,' where investors are inclined to sell profitable assets while hesitating to realize losses (Shefrin and Statman, 1985; Odean, 1998). In this context, the bid-offer setting will likely lag behind global price movements during market upturns. Specifically, on the bid side, investors may exhibit reluctance to place bids at newly elevated prices due to increased risk aversion when in profit. Conversely, on the offer side, they might set lower offers relative to global prices to facilitate selling cryptoassets. Hence, the relative quote value during market upturns is expected to be lower than during downturns. This hypothesis is visually represented in Figure 1, where a negative correlation is anticipated between relative quote values and portfolio returns.

Additionally, the opposite phenomenon, known as the 'house-money effect,' highlights that investors are more willing to take risks when their wealth exceeds certain thresholds (Thaler and Johnson, 1990). If investors behave according to this theory, they should display more aggressive quoting behavior during bullish periods, akin to "gambling with house money." This effect intersects with self-attribution and overconfidence biases, where positive portfolio returns embolden investors to assume greater risks. Therefore, a positive relationship between relative quote values and portfolio returns is expected, as depicted in Figure 1.

Another dimension of behavioral finance theory involves the tendency to underweight personal beliefs favoring market consensus (Holmes, Kallinterakis, and Ferreira, 2013). This 'herding effect' can lead to significant deviations in asset prices from their efficient levels (Bernales, Verousis, and Voukelatos, 2020) and potentially amplify market instability (Shiller, 1987; Persaud, 2000). The presence of herding suggests that bid-offer responses should swiftly align with global price movements during both upturns and downturns, driven by the fear of missing out.

In this study, our analysis of bid-offer responses is predicated solely on market reactions to global price changes, independent of whether local exchange investors possess superior information. The dynamics of bids and offers following global market movements do not necessarily reflect differential information among market participants. Thus, if the bid and offer quotes do not promptly adjust to global price levels, and the relative quote value exhibits significant variation between portfolios with positive and negative returns, it would imply the presence of behavioral biases.

In order to test the differences in bid-offer responses under different market conditions, we decided to conduct the VECM analysis on a weekly basis, starting each Sunday. This results in 109 separate estimations over the 109 weeks encompassing the period from 22 Nov 2020 to 24 Dec 2022. We assume that the implicit efficient price exhibits varying movements across weekly segments, reflecting diverse market conditions. After obtaining the VECM estimations for each week, we will compare the relative quote value with market conditions and investors' portfolio positions to test our hypotheses, as depicted in Figure 1. The detailed results of this analysis will be explored in Section IV.

## IV. Results of price discovery and behavioral biases

The first part of this section will discuss the dynamic of bid-offer spreads to understand the characteristics of trading in the crypto exchange. After this analysis, we conduct the price discovery analysis by estimating the VECM of the bid-offer quotes between the local and global exchanges. The coefficients estimation from the VECM allows us to construct the price response function and calculate the proposed metric 'relative quote value.' The estimated relative quote value will be used to test the existence of investors' behavior biases by analyzing the relationship between relative quote value and investors' portfolio returns.

Regarding the bid-offer spreads analysis, factors such as inventory costs, transaction costs, and particularly adverse selection costs are known to influence the magnitude of spreads in certain conditions. It is well-established in previous studies that adverse selection costs tend to escalate during new information announcements, overshadowing other costs (Harris, 1986; Mclnish and Wood, 1990; Foster and Viswanathan, 1993). Additionally, these studies commonly observe a U-shaped pattern in bid-offer spreads throughout the trading day, attributed to heightened trading volume and price volatility during market opening and closing times.

However, the cryptoasset market, operating 24 hours a day, presents a unique landscape devoid of traditional opening and closing periods. Our analysis of bid-offer spreads in the local exchange reveals a distinct pattern: spreads tend to be significantly elevated after midnight, peaking around 5 a.m., as illustrated in Figure 2.

Furthermore, a comparative analysis with Kraken reveals stark differences. The bid-offer spreads in the local exchange are nearly tenfold higher than those observed in Kraken. In Kraken, the spreads exhibit a relatively stable pattern throughout the day, hovering around 0.005 % for BTC, 0.013 % for ETH, and 0.061 % for XRP, as depicted in Figure 3.

**Figure 2:** Bid-offer percentage spreads in the Thai exchange

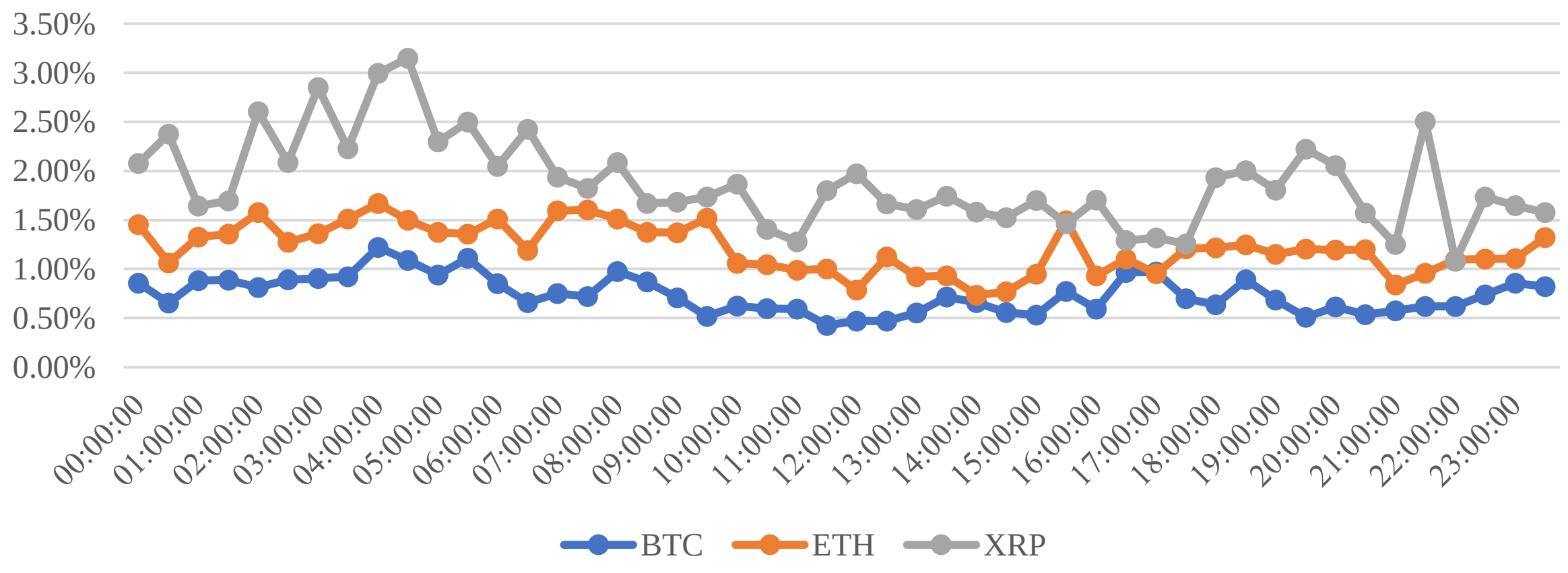

**Figure 3:** Bid-offer percentage spreads in Kraken

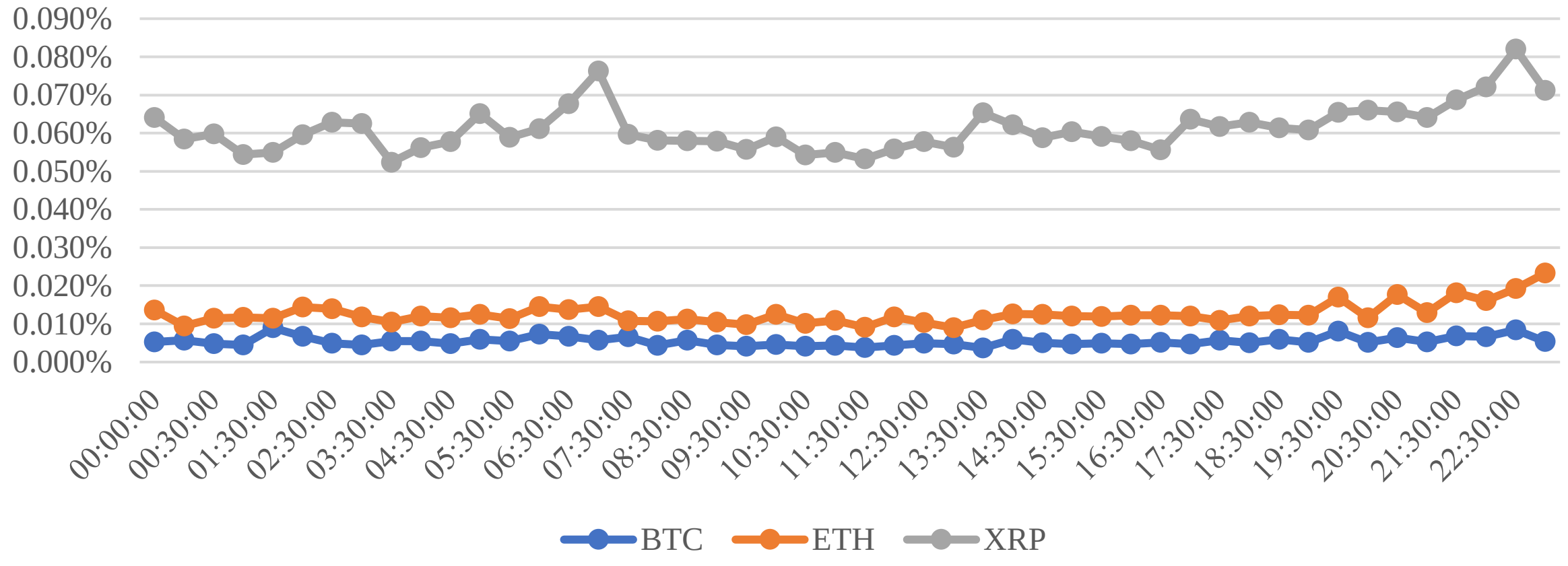

We regress bid-offer spreads on time dummies, trading volumes in each 30-minute timeframe, and volatility levels (Rogers and Satchell, 1991). We find that volatility is positively and significantly correlated with spreads (Table 3). Conversely, trading value seems to have a negligible impact on spreads, likely due to the time fixed effects.

**Table 3:** Regression results of bid-offer percentage spreads on exchange characteristics

| | Bid-offer percentage spreads of BTC | Bid-offer percentage spreads of ETH | Bid-offer percentage spreads of XRP |
|---|---|---|---|
| Ln of trading value | 0.00039*** | -0.00021*** | 0.00046*** |
| | (0.00003) | (0.00005) | (0.00007) |
| Volatility | 3.96048*** | 5.90558*** | 1.54306*** |
| | (0.15054) | (0.14605) | (0.07609) |
| Constant term | 0.00117** | 0.01408*** | 0.00935*** |
| | (0.00057) | (0.00081) | (0.00117) |
| | | | |
| R-squared | 0.063 | 0.080 | 0.049 |
| Time dummies | Yes | Yes | Yes |

Standard errors in parentheses. *** p<0.01, ** p<0.05, * p<0.1

Using the VECM estimations (Table 2), we can use the estimated parameters to formulate the expected bid-offer responses from price impulses in the global market. Figures 4 to 6 illustrate the price response of the bid-offer for a 30 % price impulse in the global market. On average, local exchange bid-offer quotes take approximately one to one-and-a-half hours to align with new global price levels. BTC exhibits the fastest response, likely due to its highest trading volume.

**Figure 4:** Bitcoin (BTC) price response in the Thai exchange from a 30 % price impulse

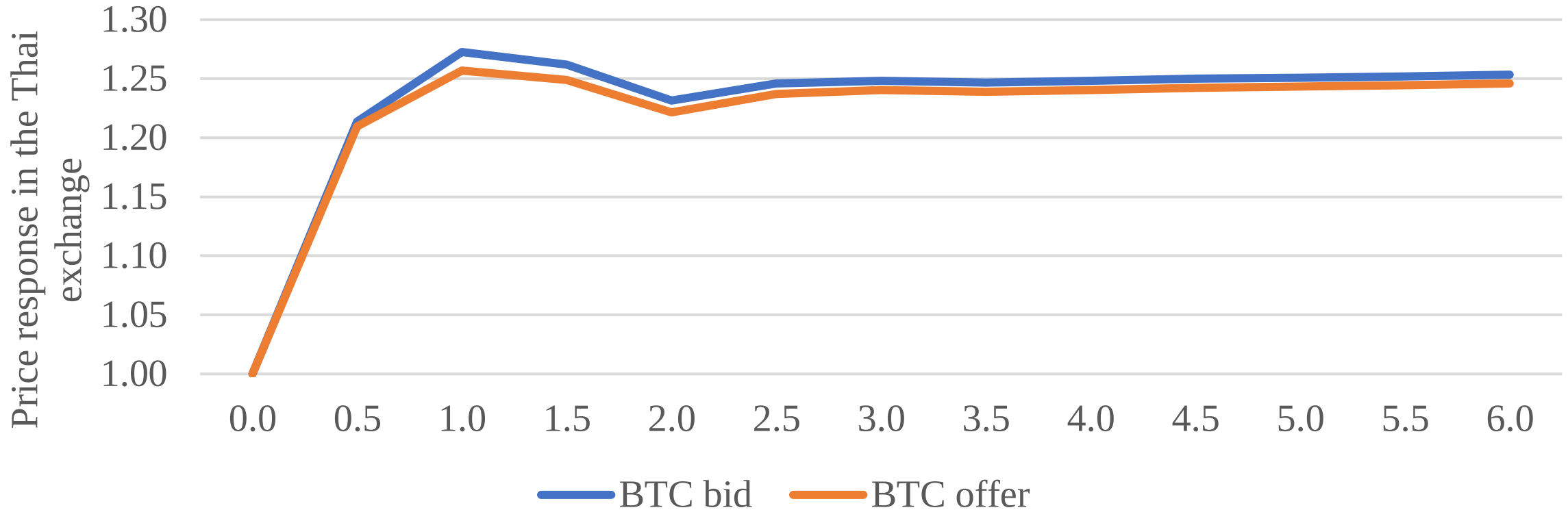

**Figure 5:** Ether (ETH) price response in the Thai exchange from a 30 % price impulse

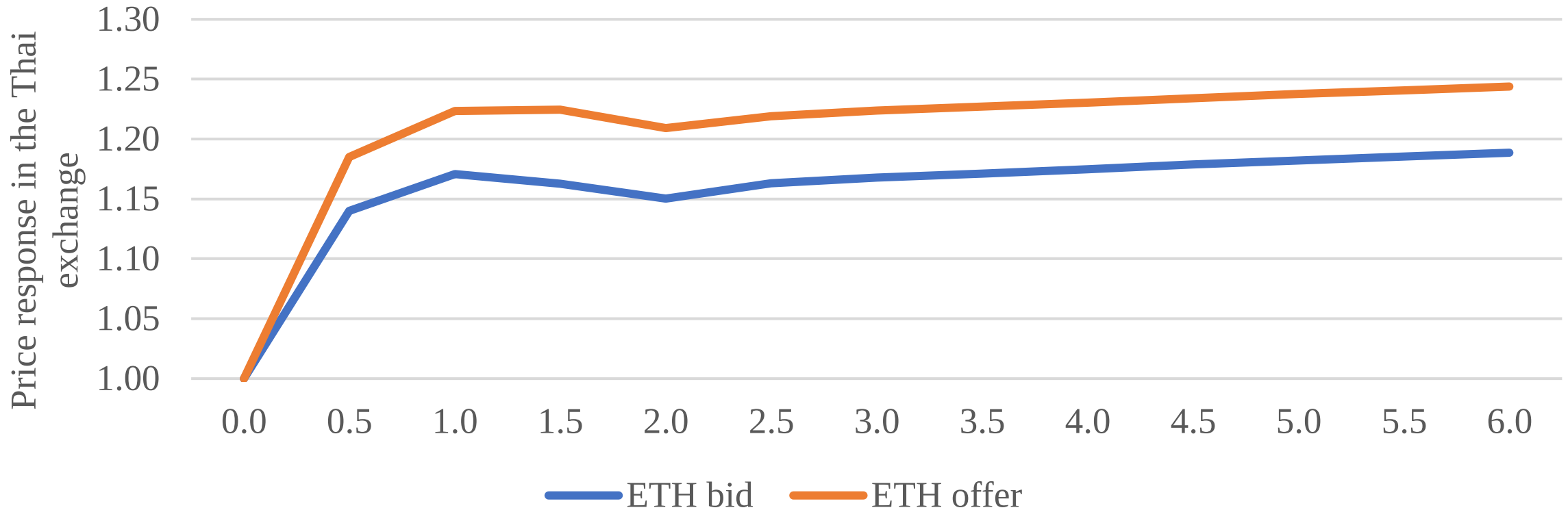

**Figure 6:** Ripple (XRP) price response in the Thai exchange from a 30 % price impulse

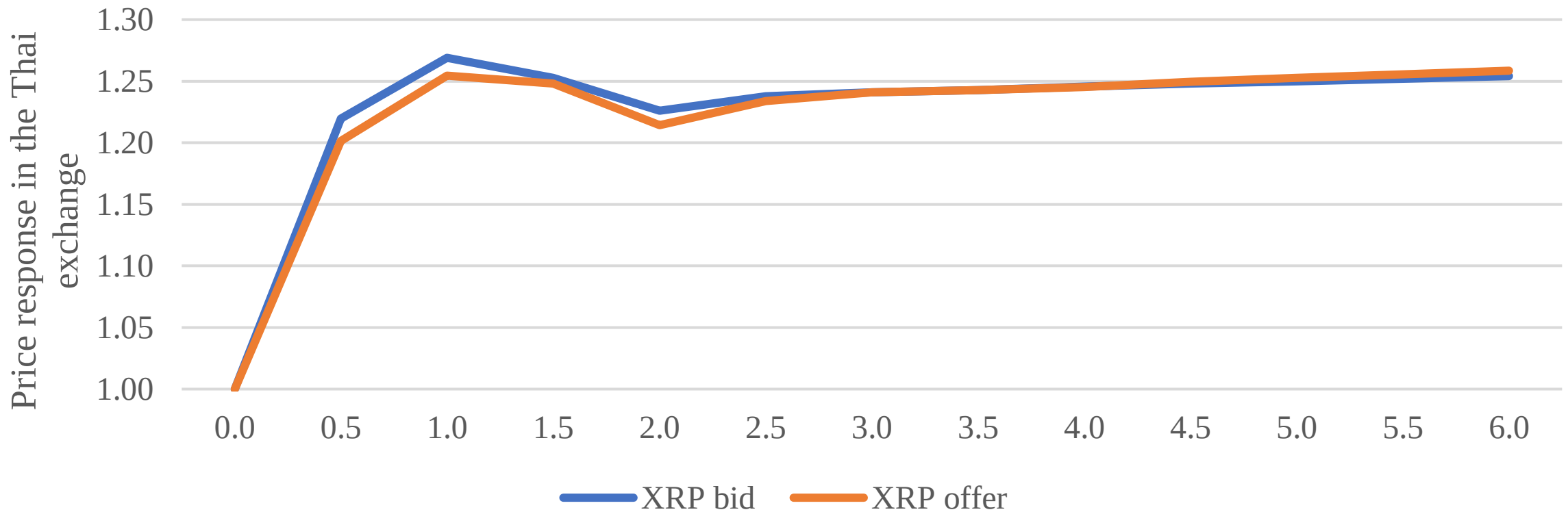

The analysis shown in Figures 4 to 6 is based on the VECM estimation of the full sample. To test how investors behave in different market conditions or portfolio gains/losses positions, we estimated the VECM model separately for each week between 22 Nov 2020 and 24 Dec 2022, resulting in 109 distinct estimations for each cryptoasset. For each week, there are 336 observations of the 30-minute bid-offer price. Table 4 summarizes the weekly trading volume and account holdings for each cryptoasset.

**Table 4:** Characteristics of accounts trading each cryptoasset

| | BTC | ETH | XRP |
|---|---|---|---|
| Weekly trading volume (unit: million baht) | | | |
| Average | 2,807.70 | 2,110.33 | 875.21 |
| SD | 1,625.20 | 1,298.18 | 1,111.64 |
| Min | 338.57 | 200.12 | 71.17 |
| Max | 8,483.33 | 6,695.16 | 7,075.35 |
| Number of accounts holding the cryptoasset | | | |
| Dec 2020 | 14,406 | 5,513 | 7,539 |
| June 2021 | 100,326 | 79,584 | 57,748 |
| Dec 2021 | 177,804 | 137,637 | 86,247 |
| June 2022 | 225,577 | 170,475 | 94,081 |
| Dec 2022 | 218,680 | 164,030 | 90,747 |

Each week, the price responses to the 30 % impulses from the estimated VECM parameters are expressed as relative quote values (outlined in Section III) at five timeframes: 30-minute, 1-hour, 1.5-hour, 2-hour, and 2.5-hour. We report the summary statistics in Table 5. The relative quote values are between 92-95 % for the 0.5-hour interval, increasing to 95-99 % after one hour and stabilizing after that.

**Table 5:** Relative quote values from a 30 % price impulse

| | Relative quote value at different timeframes | | | | |
|---|---|---|---|---|---|
| | 30-minute | 1-hour | 1.5-hour | 2-hour | 2.5-hour |
| BTC-bid | | | | | |
| Mean | 93.03% | 97.28% | 96.58% | 94.98% | 95.62% |
| SD | 2.13% | 2.43% | 2.99% | 2.94% | 2.82% |
| Min | 86.87% | 88.42% | 84.94% | 83.75% | 87.93% |
| Max | 98.46% | 105.20% | 103.93% | 100.55% | 101.56% |
| BTC-offer | | | | | |
| Mean | 92.99% | 96.83% | 96.28% | 94.78% | 95.51% |
| SD | 1.98% | 2.56% | 2.85% | 2.71% | 2.76% |
| Min | 88.61% | 87.88% | 88.28% | 86.95% | 88.89% |
| Max | 98.83% | 102.54% | 103.43% | 101.25% | 101.52% |
| ETH-bid | | | | | |
| Mean | 94.00% | 98.17% | 97.41% | 95.84% | 96.54% |
| SD | 2.19% | 2.81% | 2.98% | 2.91% | 2.91% |
| Min | 89.14% | 85.26% | 85.25% | 88.79% | 87.83% |
| Max | 100.28% | 105.14% | 103.02% | 106.45% | 106.71% |
| ETH-offer | | | | | |
| Mean | 93.87% | 97.27% | 97.04% | 95.46% | 96.28% |
| SD | 2.31% | 2.51% | 2.93% | 2.74% | 2.54% |
| Min | 88.27% | 88.23% | 85.27% | 88.08% | 89.79% |
| Max | 101.06% | 106.57% | 103.94% | 101.54% | 100.74% |
| XRP-bid | | | | | |
| Mean | 94.70% | 98.11% | 96.97% | 95.22% | 96.10% |
| SD | 2.46% | 3.20% | 3.43% | 3.36% | 3.50% |
| Min | 85.58% | 86.11% | 87.32% | 84.99% | 85.42% |
| Max | 100.90% | 107.62% | 107.03% | 103.45% | 105.46% |
| XRP-offer | | | | | |
| Mean | 93.81% | 97.21% | 96.78% | 94.85% | 95.73% |
| SD | 3.77% | 4.49% | 5.30% | 4.41% | 4.21% |
| Min | 77.48% | 77.69% | 77.50% | 77.47% | 77.84% |
| Max | 101.82% | 104.65% | 124.45% | 102.86% | 105.21% |

To test the bid-offer response variance between market upturns and downturns and to test hypotheses outlined in Figure 1, Figure 7 displays relative quote values in the 30-minute and 1-hour timeframes against cryptoasset returns. Across all three cryptoassets, a strong negative correlation exists between relative quote value and market conditions, which surprisingly persists even in the 1-hour timeframe. The negative relationship is consistent with the disposition effect.

**Figure 7:** Bid-offer responses and market conditions

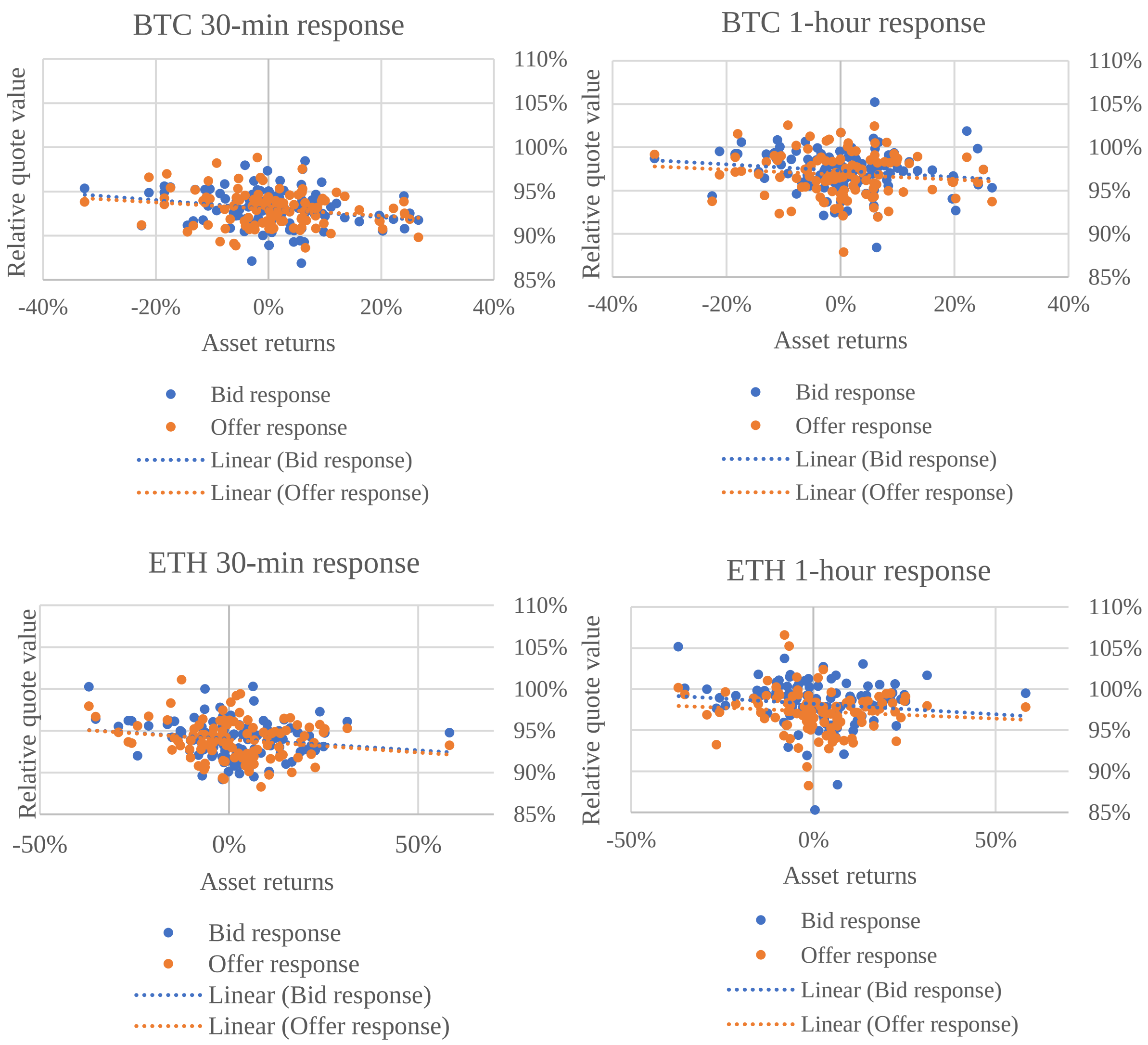

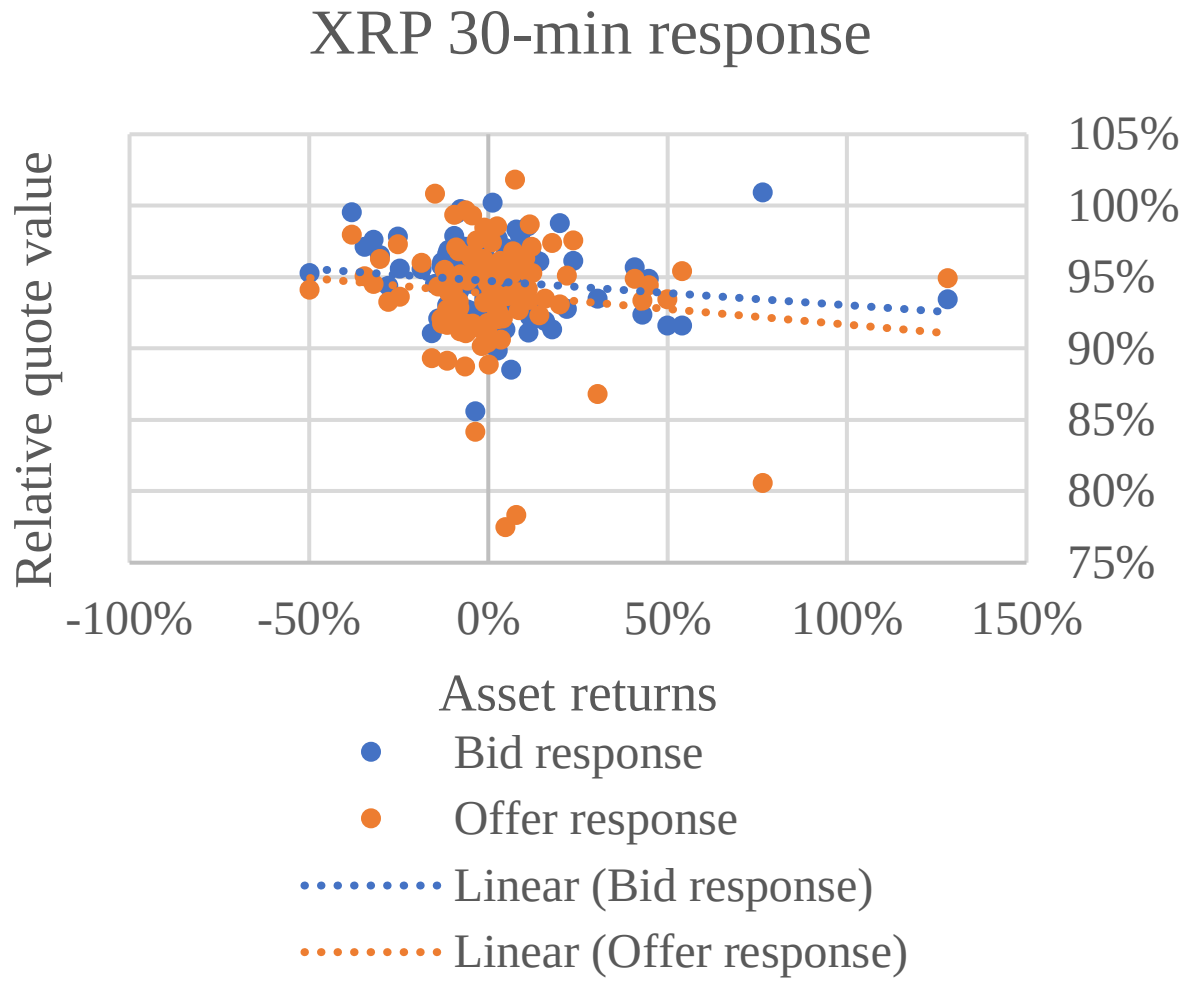

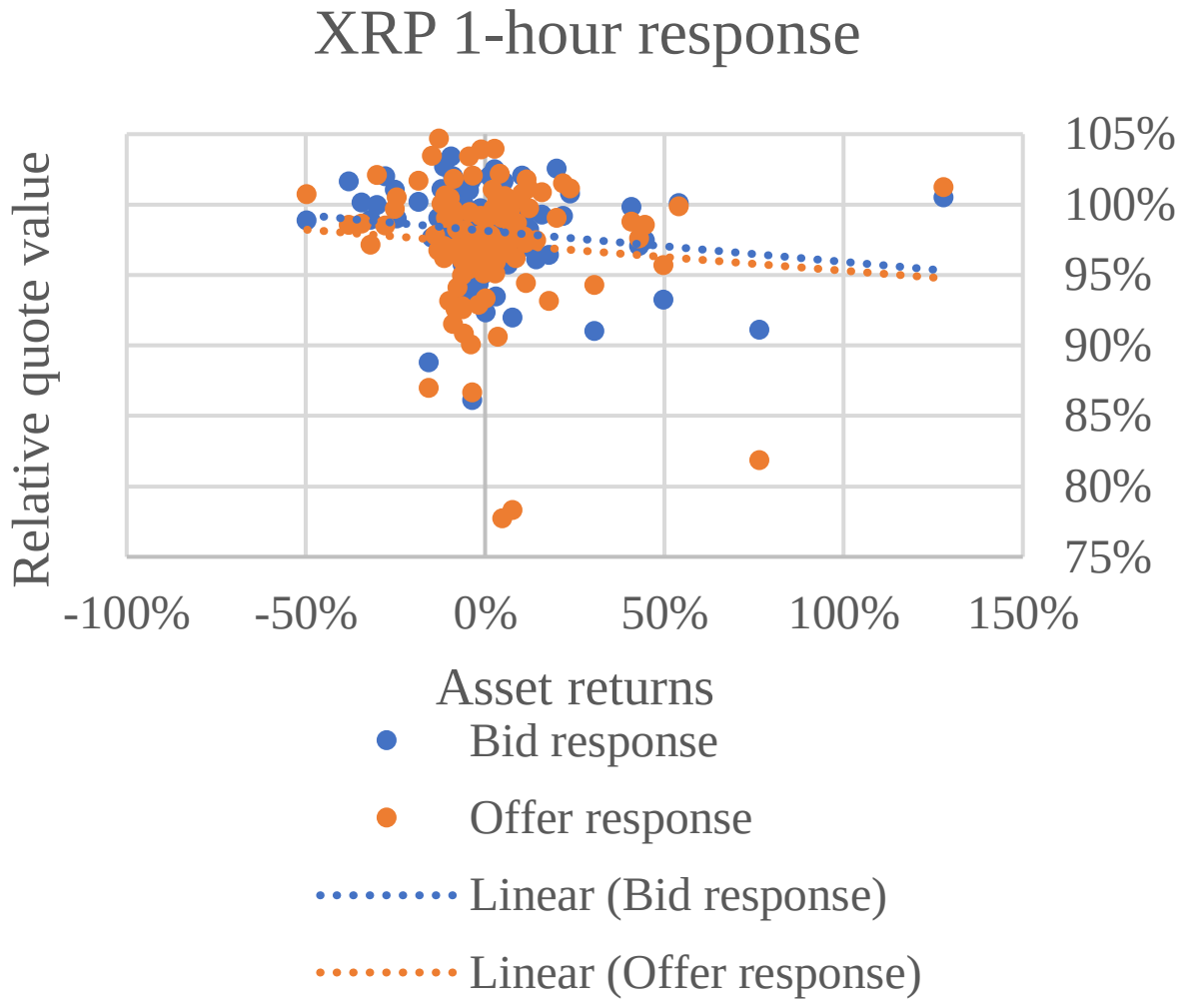

We further ascertain the negative correlation between relative quote values and the cryptoasset returns by controlling for trading value and price volatilities in a regression framework. The results in Table 6 confirm the negative correlation between relative quote value and returns: for BTC, the statistical significance is at a 5 % level, and for XRP, at a 10 % level.

**Table 6:** Regression results of relative quote values on cryptoasset returns

| | Relative quote value at 30-minute timeframe | | Relative quote value at 1-hour timeframe | |
|---|---|---|---|---|
| VARIABLES | BTC bid | BTC offer | BTC bid | BTC offer |
| Constant | 0.87873*** | 0.93531*** | 0.76860*** | 0.78650*** |
| | (0.07086) | (0.06714) | (0.08005) | (0.08563) |
| BTC weekly returns | -0.04678** | -0.03775** | -0.04104* | -0.04119* |
| | (0.01995) | (0.01891) | (0.02254) | (0.02411) |
| Ln of BTC trading value | 0.00234 | -0.00023 | 0.00946** | 0.00859** |
| | (0.00331) | (0.00314) | (0.00374) | (0.00400) |
| BTC weekly volatility | 0.11728 | -0.01680 | 0.02879 | -0.30324* |
| | (0.13252) | (0.12557) | (0.14971) | (0.16015) |
| R-squared | 0.075 | 0.038 | 0.097 | 0.065 |
| | | | | |
| VARIABLES | ETH bid | ETH offer | ETH bid | ETH offer |
| Constant | 0.84231*** | 1.17260*** | 0.85671*** | 1.09653*** |
| | (0.06744) | (0.07057) | (0.08803) | (0.08075) |
| ETH weekly returns | -0.02322 | -0.01919 | -0.02121 | -0.01096 |
| | (0.01488) | (0.01557) | (0.01943) | (0.01782) |
| Ln of ETH trading value | 0.00444 | -0.01117*** | 0.00572 | -0.00593 |
| | (0.00320) | (0.00335) | (0.00418) | (0.00383) |
| ETH weekly volatility | 0.20405** | 0.22547 | 0.21232* | 0.14081 |
| | (0.09325) | (0.09757) | (0.12170) | (0.11164) |
| R-squared | 0.121 | 0.135 | 0.086 | 0.036 |
| | | | | |
| VARIABLES | XRP bid | XRP offer | XRP bid | XRP offer |
| Constant | 0.91892*** | 0.99863*** | 0.99823*** | 1.04513*** |
| | (0.06221) | (0.09557) | (0.07942) | (0.11105) |
| XRP weekly returns | -0.02001* | -0.02221 | -0.02762* | -0.02691 |
| | (0.01138) | (0.01748) | (0.01452) | (0.02031) |
| Ln of XRP trading value | 0.00136 | -0.00312 | -0.00106 | -0.00401 |
| | (0.00315) | (0.00484) | (0.00402) | (0.00562) |
| XRP weekly volatility | 0.03131 | 0.07207 | 0.13031** | 0.22868** |
| | (0.05140) | (0.07896) | (0.06562) | (0.09175) |
| R-squared | 0.034 | 0.023 | 0.067 | 0.071 |

Standard errors in parentheses. *** p<0.01, ** p<0.05, * p<0.1

The disposition effect is related to how investors react to gains and losses in their portfolios, and asset returns are proxies for the returns on individual investors' portfolios. Fortunately, the regulatory data also includes gains and losses on individual investor's positions. Thus, we can compute the proportion of accounts with gains each week, allowing us to test the disposition effect more directly.

We repeat the analysis, replacing the cryptoasset returns with the percentage of accounts in a gain position (rather than a loss), and report the results in Figure 8 and Table 7. We find stronger negative relationships between relative quote values and the proportion of accounts with gains. Our finding indicates that retail investors are more risk-averse when their positions are profitable, preferring to realize gains, a pattern consistent even in the 1-hour interval following a price impulse.

**Figure 8:** Bid-offer responses and proportion of accounts with gains

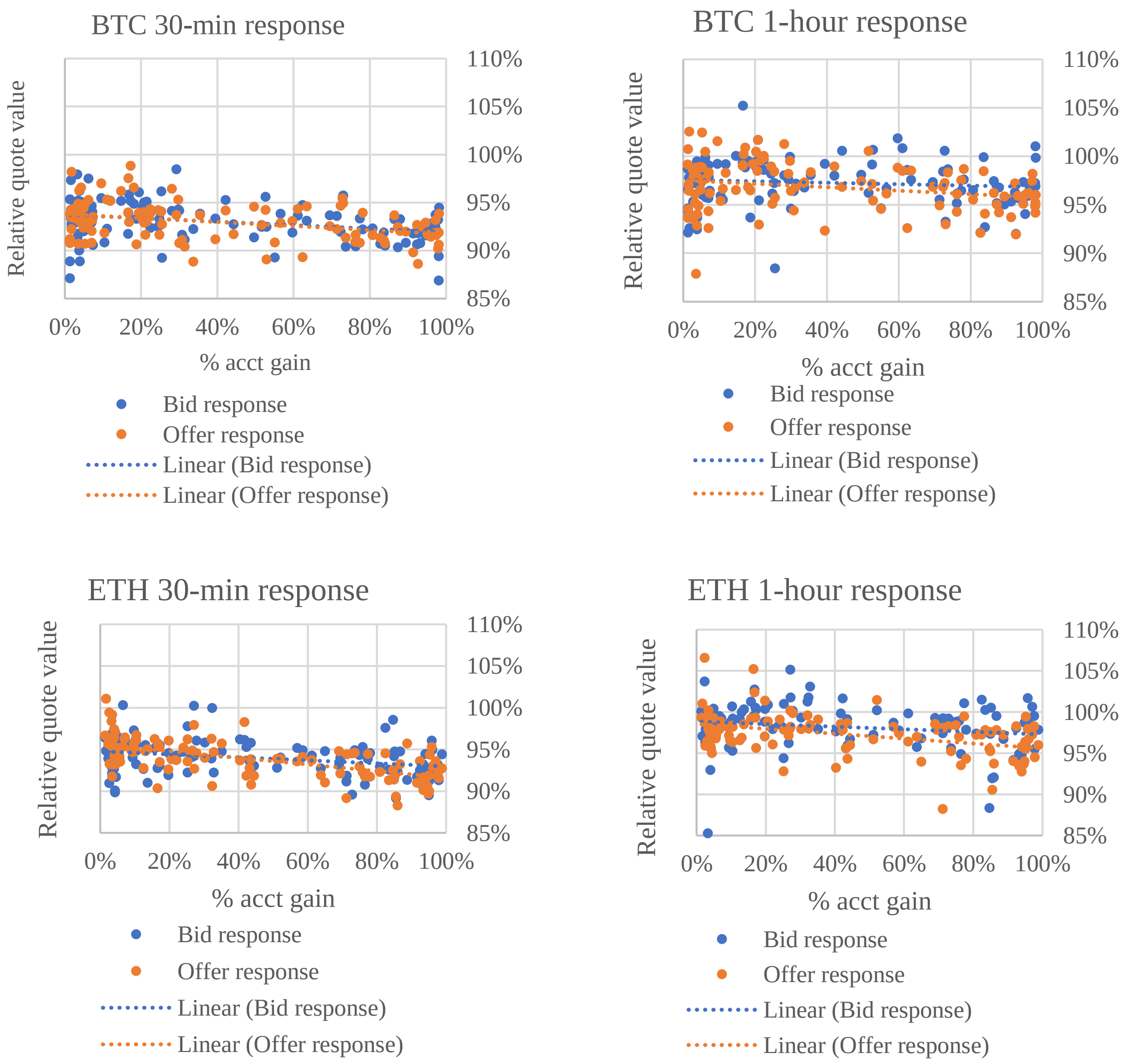

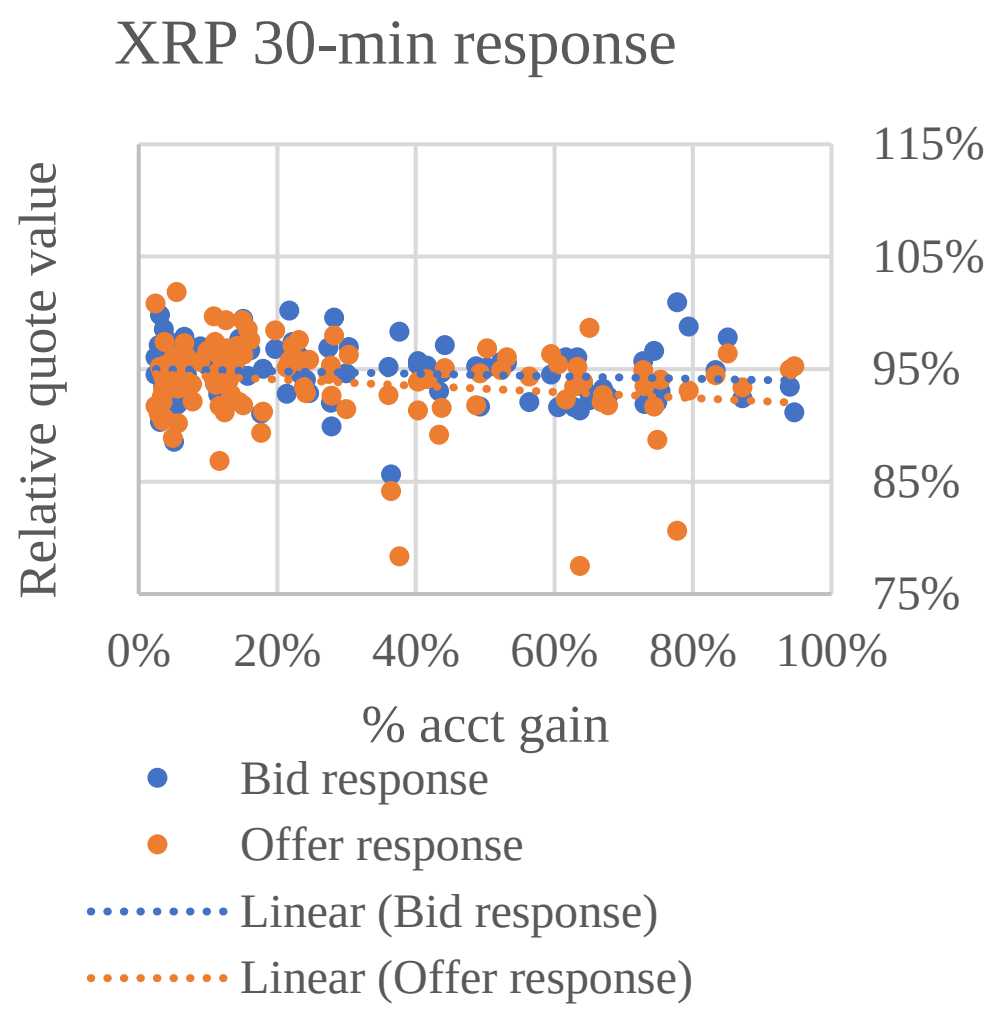

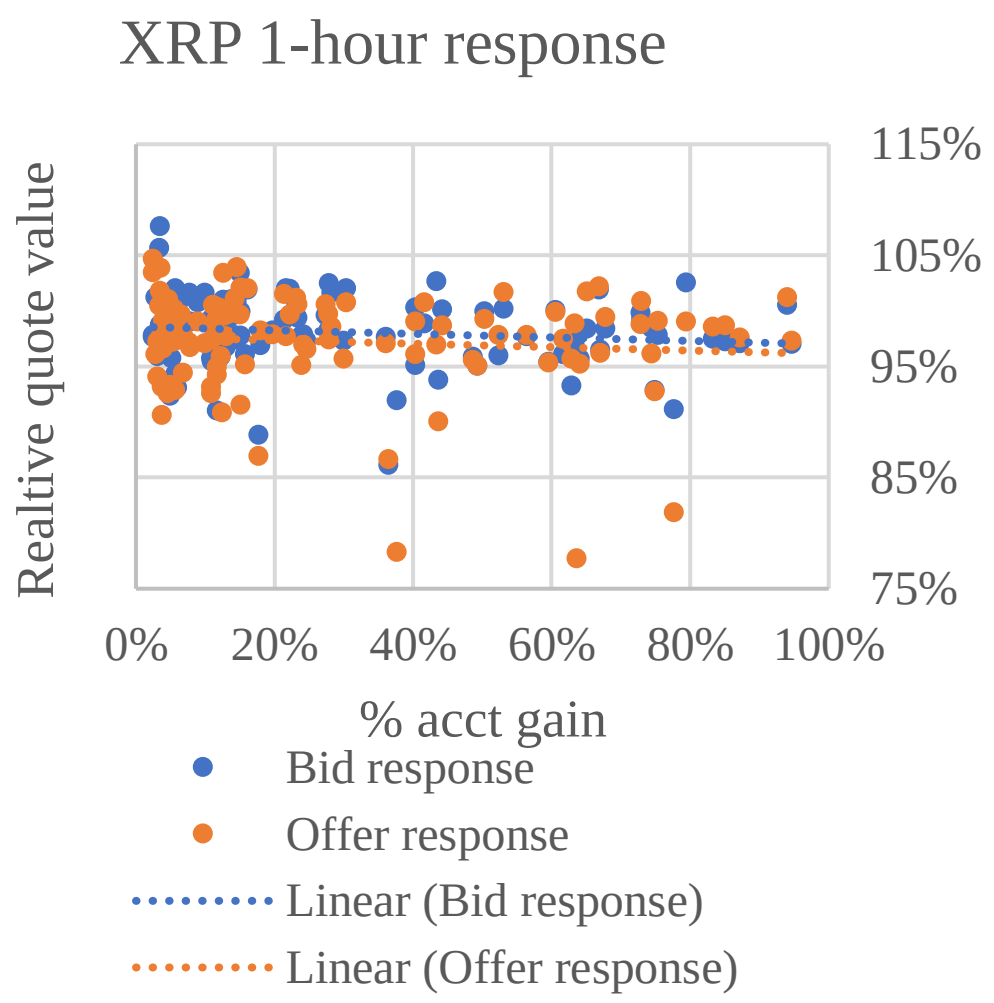

**Table 7:** Regression results of relative quote values on proportion of accounts with gains

| | Relative quote value at 30-minute timeframe | | Relative quote value at 30-minute timeframe | |
|---|---|---|---|---|
| VARIABLES | BTC bid | BTC offer | BTC bid | BTC offer |
| Constant | 0.66503*** | 0.72200*** | 0.58159*** | 0.49822*** |
| | (0.08222) | (0.07671) | (0.09575) | (0.09625) |
| % account BTC gain | -0.03320*** | -0.03245*** | -0.02905*** | -0.04310*** |
| | (0.00696) | (0.00649) | (0.00810) | (0.00815) |
| Ln of BTC trading value | 0.01289*** | 0.01029*** | 0.01870*** | 0.02281*** |
| | (0.00392) | (0.00366) | (0.00457) | (0.00459) |
| BTC weekly volatility | 0.02191 | -0.12083 | -0.05457 | -0.45349*** |
| | (0.12516) | (0.11677) | (0.14575) | (0.14651) |
| R-squared | 0.20 | 0.194 | 0.170 | 0.241 |
| Number of observations | | | | |
| VARIABLES | ETH bid | ETH offer | ETH bid | ETH offer |
| Constant | 0.76564*** | 1.03823*** | 0.78535*** | 0.98517*** |
| | (0.06737) | (0.06304) | (0.09045) | (0.07888) |
| % account ETH gain | -0.02288*** | -0.03645*** | -0.02123*** | -0.02937*** |
| | (0.00587) | (0.00550) | (0.00789) | (0.00688) |
| Ln of ETH trading value | 0.00853*** | -0.00405 | 0.00952** | -0.00004 |
| | (0.00324) | (0.00303) | (0.00435) | (0.00379) |
| ETH weekly volatility | 0.16432* | 0.13716* | 0.17500 | 0.06333 |
| | (0.08790) | (0.08226) | (0.11802) | (0.10293) |
| R-squared | 0.214 | 0.381 | 0.135 | 0.176 |
| Number of observations | | | | |
| VARIABLES | XRP bid | XRP offer | XRP bid | XRP offer |
| Constant | 0.78688*** | 0.79373*** | 0.81658*** | 0.86371*** |
| | (0.08405) | (0.12836) | (0.10704) | (0.15096) |
| % account XRP gain | -0.03486** | -0.05236** | -0.04798*** | -0.04779* |
| | (0.01366) | (0.02086) | (0.01739) | (0.02453) |
| Ln of XRP trading value | 0.00847** | 0.00788 | 0.00872 | 0.00575 |
| | (0.00437) | (0.00667) | (0.00556) | (0.00785) |
| XRP weekly volatility | 0.00848 | 0.04004 | 0.09886 | 0.19753** |
| | (0.05094) | (0.07780) | (0.06487) | (0.09149) |
| R-squared | 0.063 | 0.064 | 0.100 | 0.088 |
| Number of observations | | | | |

Standard errors in parentheses. *** p<0.01, ** p<0.05, * p<0.1

These findings support the notion that investors are more inclined to sell minor winners, underscoring the disposition effect documented by Shefrin and Statman (1985) and the concept of mental accounting (Thaler 1985). The distinct response rates between bull and bear markets suggest differing risk-return relationships, lending empirical support to Grinblatt and Han's (2005) model that momentum factors stem from the disposition effect.

Our results also suggest that buy price impacts may be smaller than sell price impacts, countering previous findings in the literature, such as Chan and Lakonishok (1993, 1995, 1997) and Holthausen, Leftwich, and Mayers (1987, 1990). This discrepancy arises from the nature of our analysis, which focuses on trading following global market trends and arbitrage activities rather than information-led activities.

Contrary to the hypotheses outlined in Section III, there is no evidence of overconfidence, self-attribution, or house-money effects in bull markets, as illustrated by Figures 7 and 8. However, the findings corroborate the literature indicating that herding behavior is more prominent in bear markets (Chiang and Zheng, 2010; Mobarek et al., 2014).

## V. Conclusion

Our in-depth analysis of quoted bid-offer orders in the Thai cryptoasset exchange provides valuable insights into investor behavior, particularly in response to global market price fluctuations. Our findings reveal behavioral biases among retail investors, predominantly the disposition effect. This bias manifests as an inclination to sell assets that are in profit. The relative quote values suggest that investors in the local exchange frequently place offers at prices lower than those in the global exchange, consistent with realizing gains on profitable positions. In addition, there is a lower tendency to place bids at high prices during market upturns, consistent with heightened risk aversion among investors with profitable positions.

This pattern of slow price response is not observed during market downturns or when investors' portfolios are experiencing losses. In these scenarios, the price response in the local exchange tends to align more closely with global price movements, supporting the efficient market hypothesis. This asymmetric behavior underscores the impact of psychological factors on investment decisions, particularly in the dynamic and evolving landscape of cryptoasset markets.

These findings corroborate the disposition effect in cryptoasset trading and highlight the divergent investor behavior under different market conditions and portfolio states. Our study contributes to the growing body of literature in behavioral finance, offering empirical evidence of behavioral biases in a market setting that defies conventional financial market structures.